\documentclass[conference,10pt]{IEEEtran}
\usepackage{graphics, subfigure, color, epsfig, psfrag, amsmath, amsfonts, amssymb, eucal, balance}
\newenvironment{sequation}{\begin{equation}\small}{\end{equation}}
\newenvironment{seqnarray}{\begin{eqnarray}\small}{\end{eqnarray}}

\newtheorem{lem}{Lemma}
\newtheorem{prop}{Proposition}

\begin{document}
\title{An Uplink Interference Analysis for Massive MIMO Systems with MRC and ZF Receivers}

\author{\authorblockN{Ning Liang$^{1}$, Wenyi Zhang$^{1}$, and Cong Shen$^{2}$}
\authorblockA{
$^1$Department of EEIS, University of Science and Technology of China\\
$^2$Qualcomm Research, San Diego, CA 92121, USA\\
Emails: {\tt liangn@mail.ustc.edu.cn, wenyizha@ustc.edu.cn, congshen@gmail.com}}}\vspace{-10pt}

\maketitle

\begin{abstract}
This paper considers an uplink cellular system, in which each base station (BS) is equipped with a large number of antennas to serve multiple single-antenna user equipments (UEs) simultaneously. Uplink training with pilot reusing is adopted to acquire the channel state information (CSI) and maximum ratio combining (MRC) or zero forcing (ZF) reception is used for handling multiuser interference. Leveraging stochastic geometry to model the spatial distribution of UEs, we analyze the statistical distributions of the interferences experienced by a typical uplink: intra-cell interference, inter-cell interference and interference due to pilot contamination.

For a practical but still large number of BS antennas, a key observation for MRC reception is that it is the intra-cell interference that accounts for the dominant portion of the total interference. In addition, the interference due to pilot contamination tends to have a much wider distribution range than the inter-cell interference when shadowing is strong, although their mean powers are roughly equal. For ZF reception, on the other hand, we observe a significant reduction of the intra-cell interference compared to MRC reception, while the inter-cell interference and the interference due to pilot contamination remains almost the same, thus demonstrating a substantial superiority over MRC reception.
\end{abstract}


\section{Introduction}
The exponential increase in demand for high data rates in cellular networks requires new technologies to boost
spectral efficiency. Recently a heightened attention has been focused on the paradigm of massive multiple input multiple output (MIMO) systems, which are envisioned as a promising key enabler for the next generation cellular network \cite{marzetta2006much} \cite{rusek2013scaling}.

It is well understood that channel state information (CSI) plays a key role in multiuser MIMO systems. A standard approach for acquiring CSI is through pilot-aided channel estimation. For massive MIMO systems, at prescribed positions of each coherence interval, all the active user equipments (UEs) in the system send pilot sequences, which are pairwise orthogonal for UEs within each cell and reused among the cells in a certain pattern; each base station (BS) correlates its received pilot signals with the known transmitted pilot sequences to obtain the estimated CSI. Due to pilot reusing, interference is inevitable among the UEs that share the same pilot sequence, and this is usually called \emph{pilot contamination} in the literature. In the asymptotic regime where the number of BS antennas grows without bound, pilot contamination remains the only source of interference \cite{marzetta2010noncooperative}. Furthermore, the interference power grows in proportion to the desired signal power, implying that the achieved spectral efficiency saturates in the asymptotic regime since the signal-to-interference ratio (SIR) is ultimately bounded.

A number of solutions have been studied aiming at breaking the aforementioned bottleneck due to pilot contamination, and they can be largely classified into two categories: subspace techniques and coordinated design. Subspace-based solutions exploit the extra degrees of freedom (DoF) provided by the large number of BS antennas to enhance estimation performance beyond linear channel estimation \cite{ngo2012evd} \cite{muller2013blind}. On the other hand, by coordinating the use of pilots or adaptively allocating and configuring pilot sequences among different UEs, solutions based on coordinated design may effectively eliminate pilot contamination under appropriate system conditions \cite{ashikhmin2012pilot} \cite{fernandes2013inter} \cite{yin2013coordinated}.

In this work, rather than solely focusing on pilot contamination in the ultimate limit of infinite BS antennas, our goal is to establish a comprehensive understanding of the various sources of interference in uplink transmission when the number of BS antennas is large but still finite. We consider a typical uplink channel in a multi-cell environment, in which the BS adopts minimum mean square error (MMSE) estimation for CSI and performs maximum ratio combining (MRC) or zero forcing (ZF) reception for each UE. Leveraging stochastic geometry to model the spatial distribution of UEs, we analyze the statistical characteristics of the interferences experienced in a typical uplink, including: intra-cell interference, inter-cell interference, and interference due to pilot contamination. The analytical results are applicable for any finite number of BS antennas, and converge to known asymptotic results as the number of BS antennas grows without bound. We have the following analytical results:
\begin{itemize}
\item For MRC reception with a practical but still large number of BS antennas, it is the intra-cell interference that accounts for the dominant portion of the total interference, and the impact of pilot contamination becomes dominant only when the ratio between the number of BS antennas $M$ and the number of available orthogonal pilot sequences $K$ is exceedingly large.
\item For ZF reception with a practical but still large number of BS antennas, on the other hand, we observe a significant reduction of the intra-cell interference compared to MRC reception, while the inter-cell interference and the interference due to pilot contamination remains almost the same.
\item In addition, for MRC reception with a practical but still large number of BS antennas in strong shadowing regime, the interference due to pilot contamination tends to have a much wider distribution range than the inter-cell interference. A similar phenomenon is also observed for ZF reception.
\item Letting $M$ and $K$ tend to infinity but maintain a fixed load factor $\kappa\triangleq K/M<1$, the mean intra-cell and inter-cell interference powers of MRC reception remain bounded away from zero, and grow in proportion with $\kappa$, and those of ZF reception grow in proportion with $\kappa/(1-\kappa)$.
\end{itemize}

The remaining part of this paper is organized as follows. Section \ref{sect:system model} outlines the system model. Section \ref{sect:training and data transmission} describes the pilot-aided uplink training scheme and derives the SINR expressions of a typical uplink for MRC and ZF receivers respectively. Then Section \ref{sect:statistic analysis} accomplishes a statistical analysis of the various interferences for the typical uplink, with the UE spatial distribution following a stochastic geometry model. Section \ref{sect:simulation result} presents numerical results to corroborate the analysis. Finally Section \ref{sect:conclusion} concludes this paper.

\section{System model}
\label{sect:system model}
We consider a cellular system in which the considered typical BS is located at the origin with its coverage area being a disk of radius $R$. The typical BS has $M$ antennas and serves $K$ uniformly distributed single-antenna UEs in its cell simultaneously. The $K$ mutually orthogonal pilots are uniquely assigned to each of the intra-cell UEs, and the UEs that reuse the $k$-th pilot outside the typical cell form a point set $\Phi_k$.To simplify notation, we let $\Phi\triangleq\cup_{k=1}^{K}\Phi_k$ denote all the UEs outside the typical cell, and $\Phi_0\triangleq\{x_1,...,x_K\}$ denote the set of UEs inside the typical cell.

We denote the channel vector between the considered BS and a UE located at $x$ by $\mathbf{g}_x$ and the channel matrix from the UEs in $\Phi_0$ to their corresponding BS by $\mathbf{G}_{0}=[\mathbf{g}_{x_1},...,\mathbf{g}_{x_K}]$. We take into account geometric path losses with both shadowing and small-scale fading, and the channel vector is thus
\begin{equation}
\mathbf{g}_x=\sqrt{\beta_x}\mathbf{h}_x,\ x\in\Phi_0\cup\Phi
\label{equ:equ_1}
\end{equation}
where $\beta_x = p_x \eta_x$ in which $p_x$ is the path loss coefficient and $\eta_x$ is the shadowing coefficient,\footnote{The shadowing is assumed to be correlated among the collocated antennas at the same BS and is thus modeled as a scalar quantity for simplicity.} and $\mathbf{h}_x \sim \mathcal{CN}(0,\mathbf{I}_{M})$ is the small-scale fading vector. We consider a block fading channel, i.e., the fading vector remains constant in each coherence interval, and changes to an independent realization in the next coherence interval. We also consider a relatively stationary scenario where $\beta_x$ remains unchanged throughout the transmission.

\section{Uplink Training and Linear Receivers}
\label{sect:training and data transmission}
\subsection{Uplink Training}
At the beginning of each coherence interval, all the UEs in the system transmit their respective pilot sequences simultaneously. Then, a typical BS correlates its received pilot signal with the known pilots, and performs a linear MMSE estimation; that is,
\begin{sequation}
\widehat{\mathbf{g}}_{y_k}=C_{y_k}\left[\mathbf{g}_{x_k}+\sum_{x\in\Phi_k}\mathbf{g}_{x}+\frac{\mathbf{z}_k}{\sqrt{\rho_p}}\right],
\label{equ:equ_2}
\end{sequation}%
where $y_k$ denotes the location of a UE that reuses the $k$-th pilot, $\rho_p$ is the uplink transmit power during the training phase, $\mathbf{z}_k\sim \mathcal{CN}(0,\mathbf{I}_M)$ represents the normalized additive white Gaussian noise (AWGN), and $C_{y_k}$ is a normalization factor taking the following form:
\begin{sequation}
C_{y_k}=\frac{\beta_{y_k}}{\beta_{x_k}+\sum_{x\in \Phi_k}\beta_x+\frac{1}{\rho_p}},\ y_k\in\{x_k\}\cup\Phi_k.
\label{equ:equ_3}
\end{sequation}%
For simplicity of notation, we further define a variable $\alpha_k$ that will be frequently used throughout this paper, i.e.,
\begin{sequation}
\alpha_k\triangleq\frac{\beta_{x_k}}{C_{x_k}}=\beta_{x_k}+\sum_{x\in\Phi_k}\beta_x+\frac{1}{\rho_p},
\label{equ:alpha_k}
\end{sequation}%
which summarizes the large-scale fading of UEs that reuse the $k$-th pilot as well as the noise term $\frac{1}{\rho_p}$. Herein we implicitly assume that the realizations of $\beta_{x}$ are perfectly known to the BS, through long-term averaging of the received signal powers or other mechanisms \cite{ashikhmin2012pilot}. 
From \eqref{equ:equ_2} and \eqref{equ:equ_3}, we have $\widehat{\mathbf{g}}_{y_k}=\frac{\beta_{y_k}}{\beta_{x_k}}\widehat{\mathbf{g}}_{x_k}$, or
\begin{equation}
\widehat{\mathbf{g}}_{y_k}=\widehat{\mathbf{G}}_0\mathbf{f}_{y_k},
\label{equ:equ_4}
\end{equation}
where $\mathbf{f}_{y_k}$ is a vector whose $k$-th element is $\frac{\beta_{y_k}}{\beta_{x_k}}$ and all the other elements are zero. Moreover, we decompose $\mathbf{g}_{y_k}$ into
\begin{equation}
\mathbf{g}_{y_k}=\widehat{\mathbf{g}}_{y_k}+\widetilde{\mathbf{g}}_{y_k},
\label{equ:equ_5}
\end{equation}
where $\widetilde{\mathbf{g}}_{y_k}$ is the estimation error vector and is independent of $\widehat{\mathbf{g}}_{y_k}$, and it holds that
\begin{eqnarray}
&\widehat{\mathbf{g}}_{y_k}\sim\mathcal{CN}(0,C_{y_k}\beta_{y_k}\mathbf{I}_M),\nonumber\\
&\widetilde{\mathbf{g}}_{y_k}\sim\mathcal{CN}(0,(1-C_{y_k})\beta_{y_k}\mathbf{I}_M).
\label{equ:equ_6}
\end{eqnarray}

\subsection{Linear Receivers}
Utilizing the estimated CSI, a typical BS constructs some linear receiver $\mathbf{W}$ to perform the multiuser reception; that is,
\begin{sequation}
\mathbf{r}=\mathbf{W}^{\text{H}}\sum_{y\in\Phi_0\cup\Phi}\mathbf{g}_{y}s_{y}+\frac{\mathbf{W}^{\text{H}}\mathbf{z}}{\sqrt{\rho_r}},
\label{equ:equ_7}
\end{sequation}%
where $s_{y}\sim \mathcal{CN}(0,1)$ denotes the data symbol transmitted by the UE located at $y$, $\rho_r$ is the uplink transmit power, and $\mathbf{z}\sim\mathcal{CN}(0,\mathbf{I}_M)$ is the normalized AWGN vector.

In this paper, we consider two widely used linear receivers MRC and ZF, i.e.,
\begin{sequation}
\mathbf{w}_k=
\begin{cases}
\sqrt{\frac{\alpha_k}{M}}\frac{\widehat{\mathbf{g}}_{x_k}}{\|\widehat{\mathbf{g}}_{x_k}\|}\ &\text{MRC},\\
\\
\widehat{\mathbf{G}}_0\left(\widehat{\mathbf{G}}^{-1}\right)_k\ &\text{ZF},
\end{cases}
\label{equ:equ_8}
\end{sequation}%
where $\widehat{\mathbf{G}}\triangleq\widehat{\mathbf{G}}_0^{\text{H}}\widehat{\mathbf{G}}_0$, $\mathbf{w}_k$ is the $k$-th column of $\mathbf{W}$, and $\mathbf{A}_{k}$ denotes the $k$-th column of matrix $\mathbf{A}$. Then, the $k$-th element of $\mathbf{r}$ can be expressed as
\begin{seqnarray}
r_{k}&\!\!\!=\!\!\!& \mathbf{w}_{k}^{\text{H}}\widehat{\mathbf{g}}_{x_k}s_{x_k}
+\sum_{y\in\Phi_0\cup\Phi\backslash\{x_k\}}\mathbf{w}_{k}^{\text{H}}\widehat{\mathbf{g}}_{y}s_{y}\nonumber\\
&&+\sum_{y\in\Phi_0\cup\Phi}\mathbf{w}_{k}^{\text{H}}\widetilde{\mathbf{g}}_{y}s_{y}
+\frac{\mathbf{w}_{k}^{\text{H}}\mathbf{z}}{\sqrt{\rho_r}}.
\label{equ:equ_9}
\end{seqnarray}%
With some calculation, the post-processing $\text{SINR}_{x_k}$ with respect to linear receiver $\mathbf{W}$ is given by
\begin{equation}
\text{SINR}_{x_k}=\frac{\left|\mathbf{w}_{k}^{\text{H}}\widehat{\mathbf{g}}_{x_k}\right|^2}
{\sum_{y}|\mathbf{w}_{k}^{\text{H}}\widehat{\mathbf{g}}_y|^2
+P_{\epsilon}\mathbf{w}_{k}^{\text{H}}\mathbf{w}_{k}},
\label{equ:equ_10}
\end{equation}
in which the summation is over $y \in \Phi_0\cup\Phi\backslash\{x_k\}$, and
\begin{sequation}
P_{\epsilon}=\sum_{y\in\Phi_0\cup\Phi}(1-C_y)\beta_y+\frac{1}{\rho_r}
\label{equ:equ_11}
\end{sequation}%
accounts for channel estimation errors and noise.

Moreover, for ZF reception, from \eqref{equ:equ_4} we have
\begin{small}
\begin{eqnarray}
|\mathbf{w}_{k}^{\text{H}}\widehat{\mathbf{g}}_y|^2&=&\mathbf{w}_{k}^{\text{H}}\widehat{\mathbf{G}}_0\mathbf{f}_y \mathbf{f}_y^{\text{H}}\widehat{\mathbf{G}}_0^{\text{H}}\mathbf{w}_{k}\nonumber\\
&=&[\widehat{\mathbf{G}}(\widehat{\mathbf{G}}^{-1})_{k}]^{\text{H}}\mathbf{f}_y \mathbf{f}_{y}^{\text{H}}[\widehat{\mathbf{G}}(\widehat{\mathbf{G}}^{-1})_{k}]\nonumber\\
&=&
\begin{cases}
\frac{\beta_y^2}{\beta_{x_k}^2}\ &\text{if}\ y\in\{x_k\}\cup\Phi_k,\\
0\ &\text{else},
\end{cases}
\label{equ:equ_12}
\end{eqnarray}
\end{small}%
and by noticing $\mathbf{w}_{k}^{\text{H}}\mathbf{w}_{k}=[\widehat{\mathbf{G}}^{-1}]_{kk}$, we can further simplify \eqref{equ:equ_10} into
\begin{equation}
\text{SINR}_{x_k}^{\text{ZF}}=\frac{\beta_{x_k}^2}{\sum_{y\in\Phi_k}\beta_y^2+\beta_{x_k}^2 P_{\epsilon}[\widehat{\mathbf{G}}^{-1}]_{kk}}.
\label{equ:equ_13}
\end{equation}

\section{Statistical Analysis of Interference}
\label{sect:statistic analysis}
In this section, we first characterize the individual interference components in a typical uplink by taking the expectation with respect to small-scale fading. After that, we specialize to a system in which the spatial distribution of UEs is described by a stochastic geometry model, and conduct a statistical analysis of the interferences. For simplicity, hereafter we ignore the effect of noise and focus on the interference-limited regime.

\subsection{Characterization of Interferences}
\label{subsect:characterization}
In this subsection, we characterize the individual interference components by taking expectation with respect to small-scale fading. We note that this exercise is reasonable since the main focus of this paper is on the comparison of the mean interferences. Moreover, averaging over fast fading provides considerable insight into the comparison between MRC and ZF, as will be shown subsequently.

The analysis for MRC reception is straightforward from \eqref{equ:equ_6} and \eqref{equ:equ_10}. Therefore we only account for the interference introduced by pilot reusing UEs as an example, given by

\begin{small}
\begin{eqnarray}
I_{\text{re}}&\!\!\!=\!\!\!&\mathbb{E}\left\{\sum_{y\in\Phi_k}|\mathbf{w}_k^{\text{H}}\mathbf{g}_y|^2\right\}\nonumber\\
&\!\!\!=\!\!\!&\mathbb{E}\left\{\sum_{y\in\Phi_k}\frac{\beta_y^2}{\beta_{x_k}^2}|\mathbf{w}_k^{\text{H}}\widehat{\mathbf{g}}_{x_k}|^2\right\}
+\mathbb{E}\left\{\sum_{y\in\Phi_k}|\mathbf{w}_k^{\text{H}}\widetilde{\mathbf{g}}_y|^2\right\}\nonumber\\
&\!\!\!=\!\!\!&\frac{M-1}{M}\sum_{y\in\Phi_k}\beta_y^2+\frac{\alpha_k}{M}\sum_{y\in\Phi_k}\beta_y.
\label{equ:equ_14}
\end{eqnarray}
\end{small}%
Note that, the first term in \eqref{equ:equ_14} is exactly the pilot contamination term that persists even when the number of BS antennas tends to infinity. As a result, we incorporate the second term in \eqref{equ:equ_14} into the inter-cell interference, and similar modification is adopted in the analysis of ZF.

In addition, we isolate the intra-cell interference $I_{\text{intra}}^{\text{MRC}}$ and the inter-cell interference $I_{\text{inter}}^{\text{MRC}}$ as follows:

\begin{small}
\begin{eqnarray}
I_{\text{intra}}^{\text{MRC}}&\!\!\!=\!\!\!&\mathbb{E}\left\{\sum_{k'\neq k}\|\mathbf{w}_k^{\text{H}}\mathbf{g}_{x_{k'}}\|^2
+\|\mathbf{w}_k^{\text{H}}\widetilde{\mathbf{g}}_{x_{k}}\|^2\right\}\nonumber\\
&\!\!\!=\!\!\!&\frac{\alpha_k}{M}\sum_{y\in\Phi_0}\beta_y-\frac{\alpha_k}{M} C_{x_k}\beta_{x_k},\\
I_{\text{inter}}^{\text{MRC}}&\!\!\!=\!\!\!&\mathbb{E}\left\{\sum_{y\in\Phi\backslash\Phi_k}
\|\mathbf{w}_k^{\text{H}}\mathbf{g}_y\|^2\right\}+\frac{\alpha_k}{M}\sum_{y\in\Phi_k}\beta_y\nonumber\\
&\!\!\!=\!\!\!&\frac{\alpha_k}{M}\sum_{y\in\Phi}\beta_y,
\end{eqnarray}
\end{small}%
noting that in (17) we incorporate the second term of \eqref{equ:equ_14}.

For ZF reception, we introduce a change of variable $\widehat{\mathbf{G}}_0\triangleq\mathbf{Z}\mathbf{D}^{1/2}$ according to \eqref{equ:equ_6}, where $\mathbf{Z}$ is a $M\times K$ matrix whose elements are drawn independently from $\mathcal{CN}(0,1)$, and $\mathbf{D}=\text{diag}\{C_{x_1}\beta_{x_1},...,C_{x_K}\beta_{x_K}\}$. Then, we have

\begin{small}
\begin{eqnarray}
\mathbb{E}\left\{\left[\widehat{\mathbf{G}}^{-1}\right]_{kk}\right\}&\!\!\!=\!\!\!&
\frac{1}{\mathbf{D}_{kk}}\mathbb{E}\left\{\left[(\mathbf{Z}^{\text{H}}\mathbf{Z})^{-1}\right]_{kk}\right\}\nonumber\\
&\!\!\!=\!\!\!&\frac{1}{K\mathbf{D}_{kk}}\mathbb{E}\left\{\text{tr}\left[(\mathbf{Z}^{\text{H}}\mathbf{Z})^{-1}\right]\right\}\nonumber\\
&\!\!\!=\!\!\!&\frac{\alpha_k}{(M-K)\beta_{x_k}^2},
\label{equ:equ_15}
\end{eqnarray}
\end{small}%
and the derivation of $I_{\text{intra}}^{\text{ZF}}$, $I_{\text{inter}}^{\text{ZF}}$ and $I_{\text{cont}}^{\text{ZF}}$ readily follows.

We summarize the results of MRC and ZF in Table \ref{tab:tab_1}. It is clear that the interferences due to pilot contamination $I_{\text{cont}}^{\text{MRC}}$ and $I_{\text{cont}}^{\text{ZF}}$ are almost the same except for a factor $(M-1)/M$ which is close to one. In addition, since $C_y$ is typically close to one for an intra-cell UE and close to zero for an outside UE, a significant reduction of the intra-cell interference and a slight attenuation of the inter-cell interference can be expected for ZF, compared to MRC. The following proposition validates the substantial suppression of intra-cell interference for ZF reception.

\begin{prop}
\label{prop:relation}
For the general system model we consider in Section \ref{sect:system model}, $I_{\text{intra}}^{\text{ZF}}$, $I_{\text{inter}}^{\text{ZF}}$ and $I_{\text{inter}}^{\text{MRC}}$ satisfy the following relationship:
\begin{equation}
I_{\text{intra}}^{\text{ZF}}<I_{\text{inter}}^{\text{ZF}}<\frac{M}{M-K}I_{\text{inter}}^{\text{MRC}}\triangleq I_{\text{intra}}^{\text{ZF,u}},
\label{equ:equ_16}
\end{equation}
\end{prop}
where $I_{\text{intra}}^{\text{ZF,u}}$ denotes the upper bound of $I_{\text{intra}}^{\text{ZF}}$ and $I_{\text{inter}}^{\text{ZF}}$.

{\it Proof:} For any $k\in\{1,...,K\}$, we have
\begin{small}
\begin{eqnarray*}
(1-C_{x_k})\beta_{x_k}&\!\!\!=\!\!\!&\frac{\beta_{x_k}\sum_{x\in\Phi_k}\beta_x}{\beta_{x_k}+\sum_{x\in\Phi_k}\beta_x}\nonumber\\
&\!\!\!<\!\!\!&\sum_{y\in\Phi_k}\frac{\beta_y[\beta_{x_k}+\sum_{x\in\Phi_k}\beta_x-\beta_y]}{\beta_{x_k}+\sum_{x\in\Phi_k}\beta_x}\nonumber\\
&\!\!\!=\!\!\!&\sum_{y\in\Phi_k}(1-C_y)\beta_y,
\label{equ:equ_17}
\end{eqnarray*}
\end{small}%
where the inequality is obtained by noticing the fact that $\sum_{x\in\Phi_k}\beta_x-\beta_y>0$ for any $y\in\Phi_k$. The remaining part of the proof is straightforward and hence omitted.

In other words, $I_{\text{inter}}^{\text{MRC}}$ behaves approximately as an upper bound of $I_{\text{intra}}^{\text{ZF}}$ and $I_{\text{inter}}^{\text{ZF}}$ as long as $K/M$ is small. Our simulation results in Section \ref{sect:simulation result} indicate that the approximation is good, especially when the effect of shadowing is not strong. Furthermore, as will be seen shortly, it is the intra-cell interference that dominates the total interference of MRC, and therefore this intra-cell interference suppression achieved by ZF is substantial. Even for moderate $M$, $\mathbf{w}_k^{\text{ZF}}$ is nearly orthogonal with the intra-cell interfering channels, while $\mathbf{w}_k^{\text{MRC}}$ is not. This explains why ZF achieves much better intra-cell interference suppression than MRC.

\begin{table}
\centering
\caption{\label{tab:tab_1}Interferences averaged over small-scale fading}
\begin{tabular}{|l|c|c|}
\hline
&ZF &MRC\\
\hline
$S$& $\beta_{x_k}^2$  &$\beta_{x_k}^2$ \\
\hline
$I_{\text{intra}}$& $\frac{\alpha_k}{M-K}\sum_{y\in\Phi_0}(1-C_y)\beta_y$ &$\frac{\alpha_k}{M}\sum_{y\in\Phi_0}\beta_y-\frac{\alpha_k}{M} C_{x_k}\beta_{x_k}$\\
\hline
$I_{\text{inter}}$&
$\frac{\alpha_k}{M-K}\sum_{y\in\Phi}(1-C_y)\beta_y$ &$\frac{\alpha_k}{M}\sum_{y\in\Phi}\beta_y$\\
\hline
$I_{\text{cont}}$& $\sum_{y\in\Phi_k}\beta_y^2$  &$\frac{M-1}{M}\sum_{y\in\Phi_k}\beta_y^2$ \\
\hline
\end{tabular}
\end{table}
\subsection{Spatial Model and Propagation Model}
\label{subsect:spatial distribution model}
Our results thus far are applicable to general system configuration. Hereafter, leveraging stochastic geometry modeling, we conduct a statistical analysis of the interferences. Specifically, we assume that $\Phi_k,k\in\{1,...,K\}$, constitute $K$ mutually independent homogeneous Poisson point processes (PPP) outside the typical cell each of intensity $\lambda$ \cite{haenggi12book}, with $\lambda\pi R^2=1$. This ensures that the density of UEs remains homogeneous throughout the system. A similar stochastic geometry model can be found in \cite{madhusudhanan2013stochastic} \cite{lin2014interplay}.

We will derive our results for general models of $\beta_x$, and for illustration purposes we will also specialize our results to a widely used simple model:
\begin{eqnarray}
p_x &=&
\begin{cases}
A_0, & |x|<d_0\\
A_0(|x|/d_0)^{-\gamma}, &|x|\geq d_0,
\end{cases}
\label{equ:equ_19}\\
\eta_x &\sim& \mathrm{LogNormal}(0, \sigma^2),
\label{eqn:shadow}
\end{eqnarray}
where $\gamma>2$ is the path loss exponent, $d_0$ is the close-in reference distance, and $A_0$ is the path loss within the close-in reference distance. For a typical outdoor scenario, $d_0$ is 100 meters and $A_0$ is $-30$dB \cite[Example 3.9]{rappaport1996wireless}\cite[Chap. 2.5]{goldsmith2005wireless}. For simplicity of notation, we define $l\triangleq d_0/R$, noting that $l < 1$ may hold in many massive MIMO applications. The shadowing coefficient $\eta_x$ follows a log-normal distribution with standard deviation $\sigma$, and is independent of $p_x$.
\subsection{Interference of MRC: Mean and Variance Analysis}
\label{subsect:average power-MRC}
To reveal the basic trends of the interferences, we evaluate their mean powers, over the underlying spatial processes and shadowing. To begin with, we introduce a lemma from \cite{haenggi12book} as follows.
\begin{lem}
\label{lem:high-order moment}
For a uniform PPP $\Phi_k$ of intensity $\lambda$, the second factorial moment measure satisfies
\begin{equation}
\mathbb{E}\left\{\sum_{x,y\in\Phi_k,x\neq y}
\beta_x\beta_y\right\}=\mathbb{E}^2\left\{\sum_{x\in\Phi_k}\beta_x\right\}.
\label{equ:high-order moment}
\end{equation}
Note that this can be generalized to any $n$-th factorial moment measure.
\end{lem}

For MRC, we have the following general result regarding the mean powers of interferences.
\begin{prop}
\label{prop:general-mean-int}
For general path loss and shadowing models, the mean interference powers of MRC are
\begin{eqnarray*}
&&\mathbb{E}[I_{\text{intra}}^{\text{MRC}}]=\frac{\lambda^2}{M}\mathbb{E}^2\{\eta\}P_i\left((K-1)P_i+KP_o\right),\\
&&\mathbb{E}[I_{\text{inter}}^{\text{MRC}}]=\frac{\lambda}{M}\mathbb{E}^2\{\eta\}\left(K\lambda P_i P_o+K\lambda P_o^2+\mathbb{E}^2\{\eta\}P_{o2}\right),\\
&&\mathbb{E}[I_{\text{cont}}^{\text{MRC}}]=\frac{(M-1)\lambda}{M}\mathbb{E}\{\eta^2\}P_{o2},\\
&&P_i\triangleq\int_{D} p_x\mathrm{d}x,\ P_o\triangleq\int_{\bar{D}} p_x\mathrm{d}x,\
P_{o2}\triangleq\int_{\bar{D}} p_x^2\mathrm{d}x,
\end{eqnarray*}
\end{prop}
where $D$ represents the coverage area of the typical cell, and $\bar{D}$ is its complementary set.

{\it Proof:} The proof follows from applying Lemma \ref{lem:high-order moment} and Campell's theorem \cite{haenggi12book} and is omitted due to space limitation.

As a case study, we have the following result.
\begin{prop}
\label{prop:mean-int}
For the path loss model \eqref{equ:equ_19} and log-normal shadowing \eqref{eqn:shadow}, the mean interference powers of MRC are
\begin{eqnarray*}
\mathbb{E}[I_{\text{intra}}^{\text{MRC}}]&\!\!\!=\!\!\!&\frac{l^4 A_0^2e^{\frac{\sigma^2}{\xi^2}}}{M(\gamma-2)^2}
\left[\gamma-2l^{\gamma-2}\right]\left[(K-1)\gamma+2l^{\gamma-2}\right],\\
\mathbb{E}[I_{\text{inter}}^{\text{MRC}}]&\!\!\!=\!\!\!&\frac{2K\gamma l^{\gamma+2}A_0^2 e^{\frac{\sigma^2}{\xi^2}}}{M(\gamma-2)^2}+
\frac{l^{2\gamma}A_0^2 e^{\frac{2\sigma^2}{\xi^2}}}{M(\gamma-1)},\\
\mathbb{E}[I_{\text{cont}}^{\text{MRC}}]&\!\!\!=\!\!\!&\frac{(M-1)l^{2\gamma}A_0^2 e^{\frac{2\sigma^2}{\xi^2}}}{M(\gamma-1)},
\end{eqnarray*}
where the unit of $\sigma$ is in dB, $\xi=10/\ln 10$, and $l\triangleq d_0/R$. Hereafter, somewhere we let $\mu\triangleq e^{\frac{\sigma^2}{\xi^2}}$ for simplicity of notation.
\end{prop}

Several observations are in order. First, comparing $\mathbb{E}[I_{\text{cont}}^{\text{MRC}}]$ and $\mathbb{E}[I_{\text{intra}}^{\text{MRC}}]$, we have $\mathbb{E}[I_{\text{cont}}^{\text{MRC}}]/\mathbb{E}[I_{\text{intra}}^{\text{MRC}}] \approx \frac{(\gamma - 2)^2}{\gamma^2 (\gamma - 1)} (M/K)\cdot l^{2\gamma - 4} \cdot e^{\sigma^2/\xi^2}$. So the interference due to pilot contamination will be dominant over the intra-cell interference only when $M$ is so large as to satisfy $M \gg K \gamma^2 l^{4 - 2\gamma} e^{-\sigma^2/\xi^2}$, but this condition is hardly met for practical system dimensions. As an example, for $d_0 = 100$m, $R=500$m, i.e., $l=1/5$, $\gamma=3.76$, and $\sigma = 8$dB, we have that $\mathbb{E}[I_{\text{cont}}^{\text{MRC}}]$ dominates $\mathbb{E}[I_{\text{intra}}^{\text{MRC}}]$ only if $M/K \gg 120$, which is unlikely to be met in practical systems since one cannot employ an extremely large number of BS antennas and be willing to serve relatively very few UEs. Second, considering $I_{\text{inter}}^{\text{MRC}}$ and $I_{\text{cont}}^{\text{MRC}}$ for the same parameter setting and $M=128$, $K=10$, we have $\mathbb{E}[I_{\text{inter}}^{\text{MRC}}]/\mathbb{E}[I_{\text{cont}}^{\text{MRC}}] \approx 5.6$ for $\sigma=3$dB and $\approx 0.31$ for $\sigma=8$dB, which indicates that $I_{\text{inter}}^{\text{MRC}}$ and $I_{\text{cont}}^{\text{MRC}}$ are typically comparable and shadowing affects $\mathbb{E}[I_{\text{cont}}^{\text{MRC}}]$ in a much more radical manner than $\mathbb{E}[I_{\text{inter}}^{\text{MRC}}]$ by an additional multiplicative factor $e^{\frac{\sigma^2}{\xi^2}}$. Moreover, it is always the intra-cell interference that accounts for the dominant part of the total interference for a practical but still large number of BS antennas.

Now, we turn to the variances of $I_{\text{inter}}^{\text{MRC}}$ and $I_{\text{cont}}^{\text{MRC}}$, which provide additional insight into the statistical difference between them. Applying Lemma \ref{lem:high-order moment}, we can obtain the following results on the variances of $I_{\text{inter}}^{\text{MRC}}$ and $I_{\text{cont}}^{\text{MRC}}$.
\begin{prop}
\label{prop:variance}
For the path loss model \eqref{equ:equ_19} and log-normal shadowing \eqref{eqn:shadow}, we have
\begin{small}
\begin{eqnarray*}
&&\text{var}[I_{\text{inter}}^{\text{MRC}}]=\\
&&\frac{l^{4\gamma}A_0^4 \mu^2}{M^2}\Bigg[\frac{\mu^6}{2\gamma-1}
+\frac{4(\gamma l^{2-\gamma}+2K)\mu^3}{(3\gamma-2)(\gamma-2)}+
\frac{(K\gamma l^{2-2\gamma}+1)\mu^2}{(\gamma-1)^2}\\
&&+\frac{4K(2\gamma l^{2-\gamma}+K\gamma l^{2-2\gamma}-1)\mu}{(\gamma-1)(\gamma-2)^2}
-\frac{4K^2(\gamma l^{2-\gamma}-2)^2}{(\gamma-2)^4}\Bigg],\\
&&\text{var}[I_{\text{cont}}^{\text{MRC}}]=\frac{(M-1)^2l^{4\gamma}A_0^4 \mu^8}{M^2(2\gamma-1)}.
\end{eqnarray*}
\end{small}%
\end{prop}

Despite of its complicated form, an immediate observation is that it is the first term that dominates $\text{var}[I_{\text{inter}}^{\text{MRC}}]$ when shadowing is strong, and therefore $\lim_{\sigma\rightarrow \infty}\text{var}[I_{\text{inter}}^{\text{MRC}}]/\text{var}[I_{\text{cont}}^{\text{MRC}}] = 1/(M-1)^2$. As a numerical study, for the same configuration as above, we have $\text{var}[I_{\text{inter}}^{\text{MRC}}]/\text{var}[I_{\text{cont}}^{\text{MRC}}]\approx 10^{-4}$ for $\sigma=8$dB. Therefore, in the strong shadowing regime, e.g., $6\text{dB}\leq\sigma\leq 8\text{dB}$, $\mathbb{E}[I_{\text{cont}}^{\text{MRC}}]$ roughly equals $\mathbb{E}[I_{\text{inter}}^{\text{MRC}}]$ but $\text{var}[I_{\text{cont}}^{\text{MRC}}]$ overwhelms $\text{var}[I_{\text{inter}}^{\text{MRC}}]$ remarkably. This phenomenon indicates that the interference due to pilot contamination tends to have a much wider distribution range than the inter-cell interference. Hence, it reveals the critical effect of shadowing on the interference due to pilot contamination from a different perspective.
\subsection{Interference of ZF: Mean Power Analysis}
\label{subsect:average power-ZF}
For ZF, closed-form expressions of $\mathbb{E}[I_{\text{intra}}^{\text{ZF}}]$ and $\mathbb{E}[I_{\text{inter}}^{\text{ZF}}]$ are available. However, due to the correlation between $(1-C_y)$ and $\beta_y$, the analytical expressions are too complicated to provide any engineering insight. Hence, we take an alternative approach by proposing a lower bound on $\mathbb{E}\{I_{\text{intra}}^{\text{ZF}}\}$, which, when combined with Proposition \ref{prop:relation}, reveals the relationship among the three interference components of ZF.
\begin{prop}
Without considering shadowing, for the path loss model given in \eqref{equ:equ_19} and the spatial distribution model in \ref{subsect:spatial distribution model}, we have the following lower and upper bounds of $\mathbb{E}[I_{\text{intra}}^{\text{ZF}}]$ and $\mathbb{E}[I_{\text{inter}}^{\text{ZF}}]$

\begin{small}
\begin{eqnarray}
\mathbb{E}[I_{\text{intra}}^{\text{ZF,l}}]&\!\!\!\!\!\!=\!\!\!\!\!\!&\frac{2K\gamma l^{\gamma+2}A_0^2}{(M-K)(\gamma-2)^2}\Bigg\{1-\frac{2l^{\gamma-2}}{K\gamma}-\frac{K-1}{2K(\gamma+2)}\cdot\nonumber\\
&&\left[2+\gamma l^{\gamma+2}\right]\left[\frac{\gamma-2}{\gamma-1}+\frac{4}{\gamma-2}\right]\Bigg\},\\
\label{equ:equ_lower}
\mathbb{E}[I_{\text{intra}}^{\text{ZF,u}}]&\!\!\!\!\!\!=\!\!\!\!\!\!&\frac{2K\gamma l^{\gamma+2}A_0^2}{(M-K)(\gamma-2)^2}+
\frac{l^{2\gamma}A_0^2}{(M-K)(\gamma-1)}.
\label{equ:equ_upper}
\end{eqnarray}
\end{small}%
\end{prop}

{\it Proof:} For any $r,t>0$, we have $\frac{rt}{r+t}>\frac{(r-t)t}{r}$. Replacing $r$ with $p_{x_{k'}}$ and $t$ with $\sum_{y\in\Phi_{k'}}p_y$, we have the following lower bound on $I_{\text{intra}}^{\text{ZF}}$, given by

\begin{small}
\begin{eqnarray*}
I_{\text{intra}}^{\text{ZF,l}}&=&\frac{p_{x_k}\sum_{x\in\Phi_k}p_x}{M-K}+\frac{p_{x_k}+\sum_{x\in\Phi_k}p_x}{M-K}\cdot\nonumber\\
&&\sum_{k'\neq k}\left(\sum_{y\in\Phi_{k'}}p_y-\frac{\left(\sum_{y\in\Phi_{k'}}p_y\right)^2}{p_{x_{k'}}}\right).
\end{eqnarray*}
\end{small}%
Then, with some algebraic manipulations we have (23). We obtain (24) by combining Proposition \ref{prop:relation} and Proposition \ref{prop:mean-int}, and we note that (24) also holds when shadowing is considered.

As an numerical study, for $d_0=100$m, $R=500$m, i.e., $l=1/5$, $\gamma=3.76$, $M=128$ and $K=10$, we have $\mathbb{E}[I_{\text{intra}}^{\text{ZF,u}}]/\mathbb{E}[I_{\text{intra}}^{\text{ZF,l}}]\approx 1.85$, thus verifying the tightness of both the lower and upper bounds for no shadowing scenario. Moreover, $\mathbb{E}[I_{\text{cont}}^{\text{ZF}}]/\mathbb{E}[I_{\text{intra}}^{\text{ZF,l}}]\approx 0.19$ for the same configuration, which indicates that the inter-cell interference is the strongest among all the interference components.
On the contrary, when shadowing is strong, e.g., $\sigma=8$dB, we have $\mathbb{E}[I_{\text{cont}}^{\text{ZF}}]/\mathbb{E}[I_{\text{intra}}^{\text{ZF,u}}]\approx 3.3$ for the same configuration. In other words, the interference due to pilot contamination tends to dominate the total interference when shadowing becomes strong. Though the quantitative relationship will be different when we change the parameters, via extensive numerical experiments, we can still conclude that the three different kinds of interference are comparable for ZF reception with a practical but still large number of BS antennas.
\subsection{The Asymptotic Regime When $\lim_{M\rightarrow\infty}\frac{K}{M}=\kappa$}
Hereafter we turn to the asymptotic regime where both $M$ and $K$ grow without bound but maintain a fixed ratio, see, e.g., \cite{zhu2014secure}. For the general system model in Section \ref{sect:system model}, we have the following result.
\begin{prop}
\label{prop:asymptotic_behaviour}
Letting $M$ and $K$ tend to infinity but maintain a fixed load factor $\kappa\triangleq \frac{K}{M}<1$, the mean intra-cell and inter-cell interference powers of MRC grow in proportion with $\kappa$, and those of ZF grow in proportion with $\kappa/(1-\kappa)$.
\end{prop}

{\it Proof:} From Table \ref{tab:tab_1}, with some manipulations we have
\begin{small}
\begin{eqnarray*}
\mathbb{E}[I_{\text{intra}}^{\text{ZF}}]\!\!=\!\!\frac{1}{M-K}A_1+\frac{K-1}{M-K}B_1,\\
\mathbb{E}[I_{\text{inter}}^{\text{ZF}}]\!\!=\!\!\frac{1}{M-K}A_2+\frac{K-1}{M-K}B_2,
\end{eqnarray*}
\end{small}%
where
\begin{equation}
A_1\!\!=\!\!\mathbb{E}\left\{\alpha_k(1-C_{x_k})\beta_{x_k}\right\},
B_1\!\!=\!\!\mathbb{E}\{\alpha_k\}\mathbb{E}\left\{(1-C_{x_k})\beta_{x_k}\right\},\nonumber
\end{equation}
\begin{equation}
A_2\!\!=\!\!\mathbb{E}\Big\{\alpha_k\!\!\!\sum_{y\in\Phi_k}\!\!(1-C_{y})\beta_{y}\Big\},
B_2\!\!=\!\!\mathbb{E}\{\alpha_k\}\mathbb{E}\Big\{\!\!\!\sum_{y\in\Phi_k}\!\!(1-C_{y})\beta_{y}\Big\}.\nonumber
\end{equation}
Letting $M$ and $K$ tend to infinity, we have
\begin{equation}
\lim_{M\rightarrow\infty}\mathbb{E}[I_{\text{intra}}^{\text{ZF}}]=\frac{\kappa}{1-\kappa}B_1,
\lim_{M\rightarrow\infty}\mathbb{E}[I_{\text{inter}}^{\text{ZF}}]=\frac{\kappa}{1-\kappa}B_2,\nonumber
\end{equation}
both of which grow in proportion with $\kappa/(1-\kappa)$. Note that $B_1$ and $B_2$ are independent of $M$ and $K$, but determined by the other system parameters as well as the spatial distribution model of UEs, according to \eqref{equ:alpha_k}. The proof for the MRC case follows analogously.
\section{Simulation Results}
\label{sect:simulation result}
\begin{figure}
\centering
\includegraphics[width=0.35\textwidth]{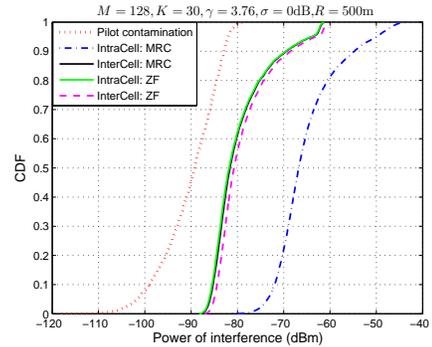}
\caption{CDF curves of the interferences: no shadowing case.}
\label{fig:figure_1}
\end{figure}

\begin{figure}
\centering
\includegraphics[width=0.35\textwidth]{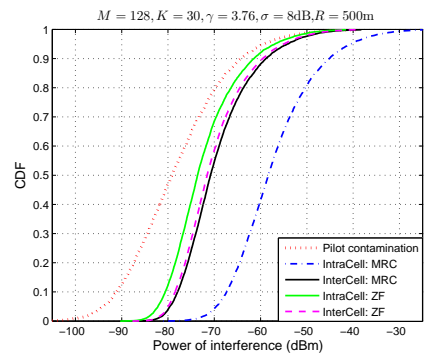}
\caption{CDF curves of the interferences: strong shadowing case.}
\label{fig:figure_2}
\end{figure}

Table \ref{tab:tab_2} summarizes the parameters used in our simulation. We choose a relatively small cell radius $R =500$m so that $l = d_0/R = 1/5$, and choose two different values of the shadowing standard deviation $\sigma = 0$dB and $8$dB. Figures \ref{fig:figure_1} and \ref{fig:figure_2} display the results of Monte Carlo simulation.

For MRC reception, both figures clearly show that the intra-cell interference is by far the dominant part of the total interference, with a 20-24dB gap above the interference due to pilot contamination (at level 50\% of CDF). The interference due to pilot contamination is even weaker than the inter-cell interference. The gaps among those interferences will be further amplified as one enlarges the cell radius (thus decreasing $l$); on the other hand, increasing the shadowing standard deviation $\sigma$ tends to reduce the gap, as can be verified from comparing Figures \ref{fig:figure_1} and \ref{fig:figure_2}.

For ZF reception, on the other hand, there exist significant reduction of the intra-cell interference compared to MRC, thus demonstrating the superiority of ZF over MRC. In addition, the figures verify that $I_{\text{inter}}^{\text{MRC}}$ provides good approximation for $I_{\text{intra}}^{\text{ZF}}$ and $I_{\text{inter}}^{\text{ZF}}$, especially when the effect of shadowing is not too strong. Moreover, both figures confirm that the three different kinds of interference become comparable, and hence suppressing the intra-cell interference alone may not be enough to harvest the gain promised by massive MIMO.

\begin{figure}
\centering
\includegraphics[width=0.35\textwidth]{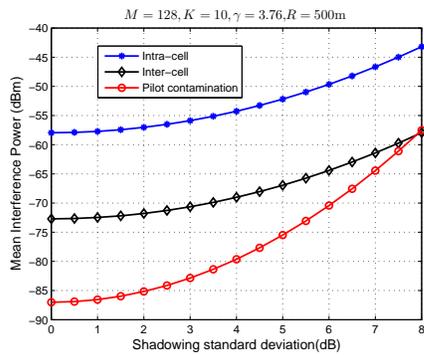}
\caption{Mean interference powers for MRC.}
\label{fig:mean_power}
\end{figure}

Figure \ref{fig:mean_power} shows the effect of shadowing on the mean interference powers by visualizing Proposition \ref{prop:mean-int}. We observe an intersection of $\mathbb{E}[I_{\text{inter}}^{\text{MRC}}]$ and $\mathbb{E}[I_{\text{cont}}^{\text{MRC}}]$ curves at $\sigma=8$dB. Moreover, shadowing boosts up $I_{\text{cont}}^{\text{MRC}}$ rapidly, thus leading to the predominance of $I_{\text{cont}}^{\text{MRC}}$ over $I_{\text{inter}}^{\text{MRC}}$ in the strong shadowing regime.

\begin{figure}
\centering
\includegraphics[width=0.35\textwidth]{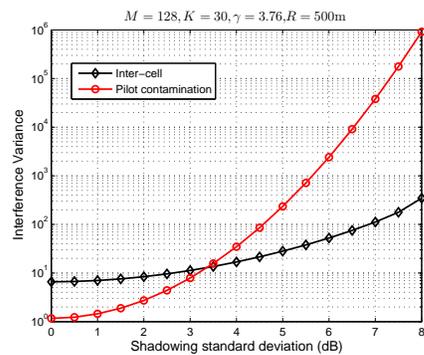}
\caption{Normalized variances of the inter-cell interference and the interference due to pilot contamination for MRC.}
\label{fig:variance}
\end{figure}

Figure \ref{fig:variance} illustrates the effect of shadowing on the normalized interference variances $\text{var}[I_{\text{inter}}^{\text{MRC}}/\mathbb{E}[I_{\text{inter}}^{\text{MRC}}]]$ and $\text{var}[I_{\text{cont}}^{\text{MRC}}/\mathbb{E}[I_{\text{cont}}^{\text{MRC}}]]$. Comparing Figure \ref{fig:mean_power} and Figure \ref{fig:variance}, it is apparent that $I_{\text{cont}}^{\text{MRC}}$ has a much wider distribution range than $I_{\text{inter}}^{\text{MRC}}$ in the strong shadowing regime, since $\mathbb{E}[I_{\text{cont}}^{\text{MRC}}]$ roughly equals $\mathbb{E}[I_{\text{inter}}^{\text{MRC}}]$ but $\text{var}[I_{\text{cont}}^{\text{MRC}}]$ overwhelms $ \text{var}[I_{\text{inter}}^{\text{MRC}}]$. This can also be verified from comparing Figure \ref{fig:figure_1} and Figure \ref{fig:figure_2}, where $I_{\text{cont}}^{\text{MRC}}$ has a much longer tail at the near zero end of the CDF curves.
\begin{table}
\centering
\caption{\label{tab:tab_2}Simulation parameters}
\begin{tabular}{|l|l|}
\hline
Parameter &Description\\
\hline
Cell radius $R$  & 500m \\
Close-in reference distance $d_0$ & 100m\\
Close-in path loss $A_0$ & -30dB\\
Pathloss exponent $\gamma$ &3.76\\
Shadowing standard deviation $\sigma$ & 0dB$\sim$8dB\\
Number of BS antennas $M$ &128\\
Number of pilots $K$ &30\\
Normalized transmit power $\rho_p$ &0dB\\
\hline
\end{tabular}
\end{table}
\section{Conclusion}
\label{sect:conclusion}
We analyzed the interferences in an uplink multi-cell massive MIMO system, leveraging stochastic geometry to model the spatial distribution of the UEs. For a practical number of BS antennas, we concluded that it is the intra-cell interference that accounts for the dominant portion of the total interference of MRC, and that a significant reduction of the intra-cell interference is achieved by ZF. It is expected that the conclusion is fairly robust against propagation models and UE distributions. An interesting problem for future work is to analyze the interferences under receivers beyond MRC and ZF, for example, regularized zero-forcing or MMSE, which is expected to balance the different interference components by adjusting the regularizing factor.

\section*{Acknowledgement}
The research has been supported by National High Technology Research and Development Program of China (863 Program) through grant 2014AA01A702, National Natural Science Foundation of China through grant 61379003, and 100 Talents Program of Chinese Academy of Sciences.


\end{document}